\documentclass[twocolumn,prl,superscriptaddress,nofootinbib,amsmath,amssymb, aps, floatfix, longbibliography]{revtex4-1}

\usepackage{graphicx}
\usepackage{bm}
\usepackage{amsmath}
\usepackage{xcolor}
\usepackage[colorlinks=true,citecolor=blue,linkcolor=magenta]{hyperref}
\usepackage[english]{babel}
\usepackage[ansinew]{inputenc}
\usepackage{boldline,multirow,xcolor,colortbl}
\usepackage{times}
\usepackage{graphicx}
\usepackage{graphics}
\usepackage{amsmath}
\usepackage{amsfonts}
\usepackage{amssymb}
\usepackage{amsbsy}
\usepackage{dsfont}
\usepackage{epstopdf}
\usepackage{makeidx}
\usepackage{subfigure}
\usepackage{color} 
\usepackage{pgf}
\usepackage{bm}
\usepackage{tikz} 
\usepackage[normalem]{ulem}
\usepackage[symbol]{footmisc}

\newcommand{\ben}{\begin{eqnarray}}
\newcommand{\een}{\end{eqnarray}}

\newcommand{\be}{\begin{equation}}
\newcommand{\ee}{\end{equation}}

\begin{document}

\title{Ultrafast Nonequilibrium Dynamics in Two-dimensional Quantum Spin-Hall Materials}
\author{Rajesh K. Malla}
\affiliation{Theoretical Division, MS B262, Los Alamos National Laboratory, Los Alamos, New Mexico 87545, USA}
\affiliation{Center for Nonlinear Studies, MS B258, Los Alamos National Laboratory, Los Alamos, New Mexico 87545, USA}
\author{Dasol Kim}
\affiliation{Department of Physics and Center for Attosecond Science and Technology, POSTECH, 7 Pohang 37673, South Korea}
\affiliation{Max Planck POSTECH/KOREA Research Initiative, Pohang 37673, South Korea}
\author{Dong Eon Kim}
\affiliation{Department of Physics and Center for Attosecond Science and Technology, POSTECH, 7 Pohang 37673, South Korea}
\affiliation{Max Planck POSTECH/KOREA Research Initiative, Pohang 37673, South Korea}
\author{Alexis Chac\'{o}n}
\affiliation{Department of Physics and Center for Attosecond Science and Technology, POSTECH, 7 Pohang 37673, South Korea}
\affiliation{Max Planck POSTECH/KOREA Research Initiative, Pohang 37673, South Korea}
\author{Wilton J. M. Kort-Kamp}
\email{kortkamp@lanl.gov}
\affiliation{Theoretical Division, MS B262, Los Alamos National Laboratory, Los Alamos, New Mexico 87545, USA}

\date{\today}

\begin{abstract}
We develop the theoretical framework of non-equilibrium ultrafast  photonics in monolayer quantum spin-Hall insulators supporting a multitude of topological states. In these materials, ubiquitous strong light-matter interactions in the femtosecond scale lead to non-adiabatic quantum dynamics, resulting in topology-dependent nonlinear optoelectronic transport phenomena. We investigate the mechanism driving topological Dirac fermions interacting with strong ultrashort light pulses and  uncover various experimentally accessible physical quantities that encode  fingerprints of the quantum material's topological  electronic state from the high harmonic generated spectrum. Our work sets the theoretical cornerstones to realize the full  potential of time-resolved harmonic spectroscopy for identifying topological invariants in  two-dimensional quantum spin-Hall solid state systems.
\end{abstract}

\maketitle

Two-dimensional quantum spin-Hall insulators (QSHI) are atomically thin materials that support counter-propagating helical metallic spin edge states with zero net electronic conductance \cite{bernevig2006,Hasan2010,Sci1,Sci2,Natphys}. This new quantum state of matter is protected by time-reversal symmetry and the robustness of its chiral spin currents against disorder, perturbations, and dissipation could potentially serve as a router for coherent information transport across the nodes of quantum networks, enabling new technologies in quantum information sciences \cite{Manchon2015, Tokura2017}. Kane and Mele suggested that spin-orbit coupling in 2D materials produce a gap in the energy band-structure that ultimately results in a QSHI state \cite{KM1,KM2}. Although the successful exfoliation of  graphene \cite{Novoselov2004} paved the way for a potential experimental demonstration of various Hall effects in monolayer systems \cite{Hall1,Hall2,Hall3,Hall4,Hall5,Hall6,Hall7,Hall8}, the realization of QSHI states in graphene remains elusive due to its minimal spin-orbit coupling \cite{smallSO}. Various other two-dimensional materials with stronger spin-orbit coupling \cite{Gomez2016, Molle:2017aa, Mannix:2017aa}  that can serve as quantum spin Hall insulators have been proposed, such as antiferromagnetic manganese chalcogenophosphates \cite{Li} and perovskites \cite{Liang}, and realized experimentally, including  silicene \cite{Silicene}, germanene \cite{Germanene}, stanene \cite{Stanene}, and plumbene \cite{Plumbene}, and recently developed jacutingaite materials \cite{J1,J2,J3,J4}. Remarkably, many of these materials are predicted to support a multitude of topological phases that can be controlled via external interactions and fields \cite{EzawaJapanese, KortKamp2017, Wiltonnature}, providing an all-in-one material platform for on-demand multi-optoelectronic functionalities.

Nonlinear optical spectroscopy has been a go-to method to probe quantum systems with discrete energy levels, e.g., in atomic and molecular systems \cite{AMO1,AMO2,AMO3,AMO4,AMO5,AMO6}. It has also been extended to solid-state systems, in which case an intense laser pulse excites charge carriers to highly non-equilibrium states and the corresponding  spectra, resulting from high-harmonic generation (HHG), serves as a tool to examine material properties, including electronic structures and crystal symmetries \cite{Ciappina_2017,Basov2017,Liu2017a,You2017a,Ghimire2018,Vampa2019,Kovalev2020}. Single-atomic-layer solids have become an attractive platform to elucidate the underlying mechanisms governing nonlinear responses and high-harmonic generation since they do not suffer from phase-matching condition effects \cite{Gierz2013, Yoshikawa2017, Hafez2018, Zhang2018a, Baudisch2018, Hafez2020a}.  Recently, it has been proposed that ultrafast spectroscopy could be used to study chiral Hall states of quantum materials \cite{Morimoto2016a,Reimann2018,Luu2018,Avetissian2018a}. Also, it has been shown that sub-gap harmonic generation is significantly enhanced in topological phases of finite-size one-dimensional chains of nanoparticles \cite{RSeq, HHG4, Drueeke2019} and Haldane nanoribbons \cite{BauerHaldane}. Moreover,  circular-dichroism and helicity of the emitted harmonics could be employed for sensing the topology of the electronic band-structure of 2D Chern insulators \cite{HHG5, HHG6, Chacon2020}. Even more recently, non-integer high-harmonics originating from surface states in three-dimensional bismuth-telluride insulators have been experimentally observed \cite{nonint}. 

Dynamic strong-field interactions and the use of ultrafast nonlinear spectroscopy to investigate phase transitions involving quantum states with distinct topology in QSHI monolayers remain largely uncharted to date. Here, we bridge this knowledge gap by developing the theoretical cornerstones of the non-equilibrium dynamics of Dirac-like topological fermions in two-dimensional quantum spin Hall insulators, and use it to study topological phase transition fingerprints beyond linear response regime. We unveil the physical mechanisms driving ultrafast hot electron population excitation and relaxation in a multitude of topologically protected quantum states supported by two-dimensional QSHI. We show that competing intraband and interband transitions govern the trade-off between nonlinear Hall and longitudinal currents, leading to emerging ultrafast  effects  due to the back-action of charge carriers in the optical field. We demonstrate non-adiabatic quantum electronic transport in the strong field interaction regime and discover several physical quantities, both in time and frequency domains,  that can serve as metrological probes of the topology of the energy band structure in these materials. Our results build the  theoretical tools in ultrafast photonics to understand non-equilibrium processes in quantum topological systems and are of utmost relevance to  materials science, photonics, metrology, and spintronics, among other areas. 
\begin{figure}
\includegraphics[width=1.0\linewidth]{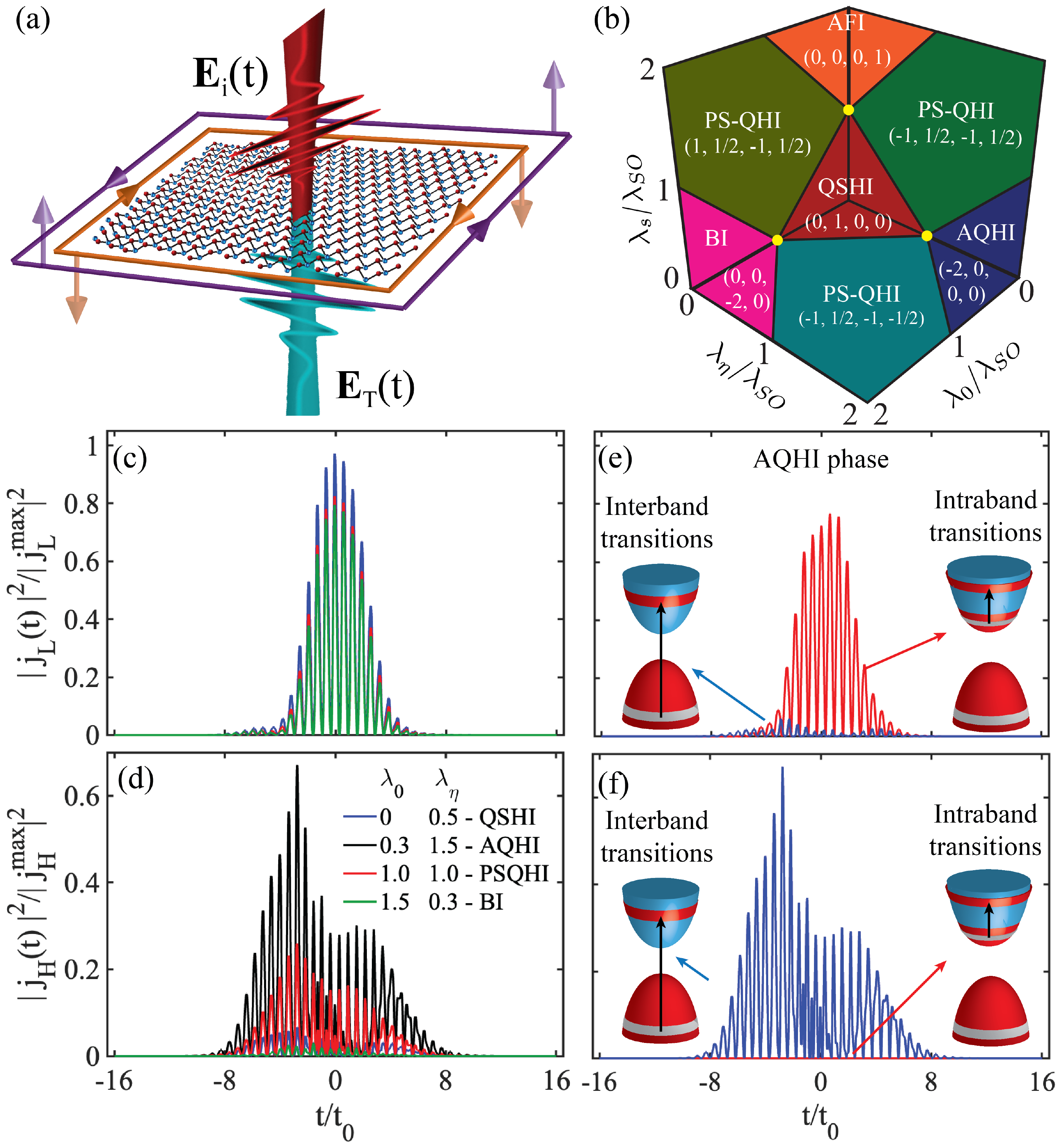}
\caption{{\bf Driven two-dimensional quantum spin-Hall insulators.} (a) Srong-field physics in a generalized Kane-Mele quantum spin-Hall monolayer interacting with an ultrashort probe optical pulse. (b) Electronic phases supported by the system include quantum spin Hall insulator (QSHI), anomalous quantum Hall insulator (AQHI), band insulator (BI), antiferromagnetic insulator (AFI), and polarized-spin quantum Hall insulator (PS-QHI) \cite{EzawaJapanese, KortKamp2017, Wiltonnature}. Their corresponding Chern, spin Chern, valley Chern, and spin-valley Chern numbers, respectively, are shown in parenthesis. Black lines mark phase transition boundaries identifying the closing of a single mass gap. Bright points represent quadruple-phase boundaries with two vanishing gaps. (c) Longitudinal  and (d) Hall currents for selected phases, normalized by the maximum achieved current.  The contributions from  intraband and interband transitions to $j_L(t)$ and $j_H(t)$ are shown in (e) and (f), respectively, for the AQHI phase. The parameters of the incident laser are $ev_F A_0/\lambda_{SO} = 5 $, $\hbar\omega_0/\lambda_{SO} = 5$, and $\tau = 8t_0$, where  $t_0 = \hbar/\lambda_{SO}$. We assume an electronic relaxation rate of $\hbar \Gamma/\lambda_{SO} = 0.05$ and an equilibrium Fermi energy $E_F = 0$.}
\label{F1}
\end{figure}

Let us consider a generalized two-dimensional quantum spin-Hall insulator interacting with an ultrashort  optical pulse, as shown in Fig. \ref{F1}a.   For concreteness, we consider that the energy band structured of the monolayer is described by a generalized Kane-Mele Hamiltonian \cite{mallasp, mallaprb} ${H}_s^{\eta}({\bf k})= \hbar v_F(\eta k_x \tau_x +k_y \tau_y) + \Delta_s^{\eta} \tau_z$, which captures all the topological properties and dynamics of the full tight-binding model in the low-energy regime \cite{EzawaJapanese}. Here, ${\textbf{p}}=  \hbar {\bf k} = \hbar(k_x,k_y)$ is the particle momentum, $v_F$ is the Fermi velocity, $\tau_{i}$ are the sub-lattice pseudo-spin Pauli matrices, and $\eta, s = \pm 1$ are valley and spin indexes. The generalized mass term $\Delta_s^{\eta}=\eta s \lambda_{SO}-\lambda_{0}-\eta \lambda_{\eta}+s\lambda_{s}$ is the half energy band gap for a particular Dirac cone. It is determined by the intrinsic spin-orbit coupling $\lambda_{SO}$ as well as by three knobs $\lambda_0, \ \lambda_\eta, \ \lambda_s$ representing interactions with external systems or fields  \cite{EzawaJapanese}. The term $\lambda_0$ describes, e.g., the staggered sub-lattice potential induced by a static electric field applied normal to a monoatomic monolayer of the graphene family \cite{nicol2012, WiltonNonlocal}. The component $\lambda_{\eta}$ corresponds to a second-neighbor hopping that arises due to the coupling with a high-frequency off-resonant circularly polarized laser that can induce an anomalous quantum Hall phase \cite{oka2009, Ezawaphoto, Mclver2020}. The final term $\lambda_s$ depicts the anti-ferromagnetic exchange interaction that emerges due to interaction of the two-dimensional material with a substrate \cite{stagger}.  On demand manipulation of $\lambda_0,\lambda_{\eta},$ and $\lambda_s$ enables a wealth of topological  phases and transitions, as shown in Fig.~\ref{F1}b. 

The impinging light drives the monolayer out-of-equilibrium and photoexcites a non-thermal free carrier density. Following electron relaxation and electron-hole recombination mechanisms lead to the emission of harmonics of the incident field, which encode signatures of the material's energy band structure. The transmitted optical pulse $\textbf{E}_T(t)$ follows from Maxwell's equations subjected to boundary conditions at the monolayer considering the induced longitudinal and Hall surface currents $\textbf{j}(t) = j_L(t)\hat{\textbf{x}}+j_H(t)\hat{\textbf{y}}$. At normal incidence $\textbf{E}_T(t) = -\partial_t{\bm A}(t) - \mu_0 c \textbf{j}(t)/2$, where  ${\bm A}(t)= A_0 e^{-(4\log 2) t^2/{\tau}^2}\cos (\omega_0 t) {\hat {\bf x}}$ is the vector potential of the linearly polarized incident field. The quantum dynamics of generalized Kane-Mele quantum spin-Hall insulators, and their topology footprints in the optical field via $\textbf{j}(t)$, is obtained from the above Hamiltonian through a minimal fermion-light coupling substitution \cite{sipe1993} $\hbar {\bm k} \rightarrow  \boldsymbol{\Pi}_{\textbf{k}}(t) =  {\bm p}-e {\bm A}(t)$. The time dependence of the Hamiltonian through ${\bm A}(t)$ prevents analytical solutions to the Dirac equation beyond the weak coupling regime. Non-perturbative results under extreme nonlinear optical conditions can be accomplished by extending the  formalism of Refs. \cite{Ishikawa2010, Ishikawa2013} to massive topological fermions . It consists in writing the electronic spinor $ |\psi_{{\bm k}}^{\eta,s}(t)\rangle$ as a linear combination of instantaneous eigenstates of ${H}_s^{\eta}(\boldsymbol{\Pi}_{\textbf{k}}(t) )$, which  enables one to derive coupled Dirac-Bloch differential equations for the population difference and interband coherence. Interestingly, these equations can be made SOC-invariant by expressing all physical quantities in units of $\lambda_{SO}$, making their solutions and the ensuing electronic currents universal to any Kane-Mele QSHI material (Supplemental Material \cite{SI}).
\begin{figure}
\includegraphics[width=1\linewidth]{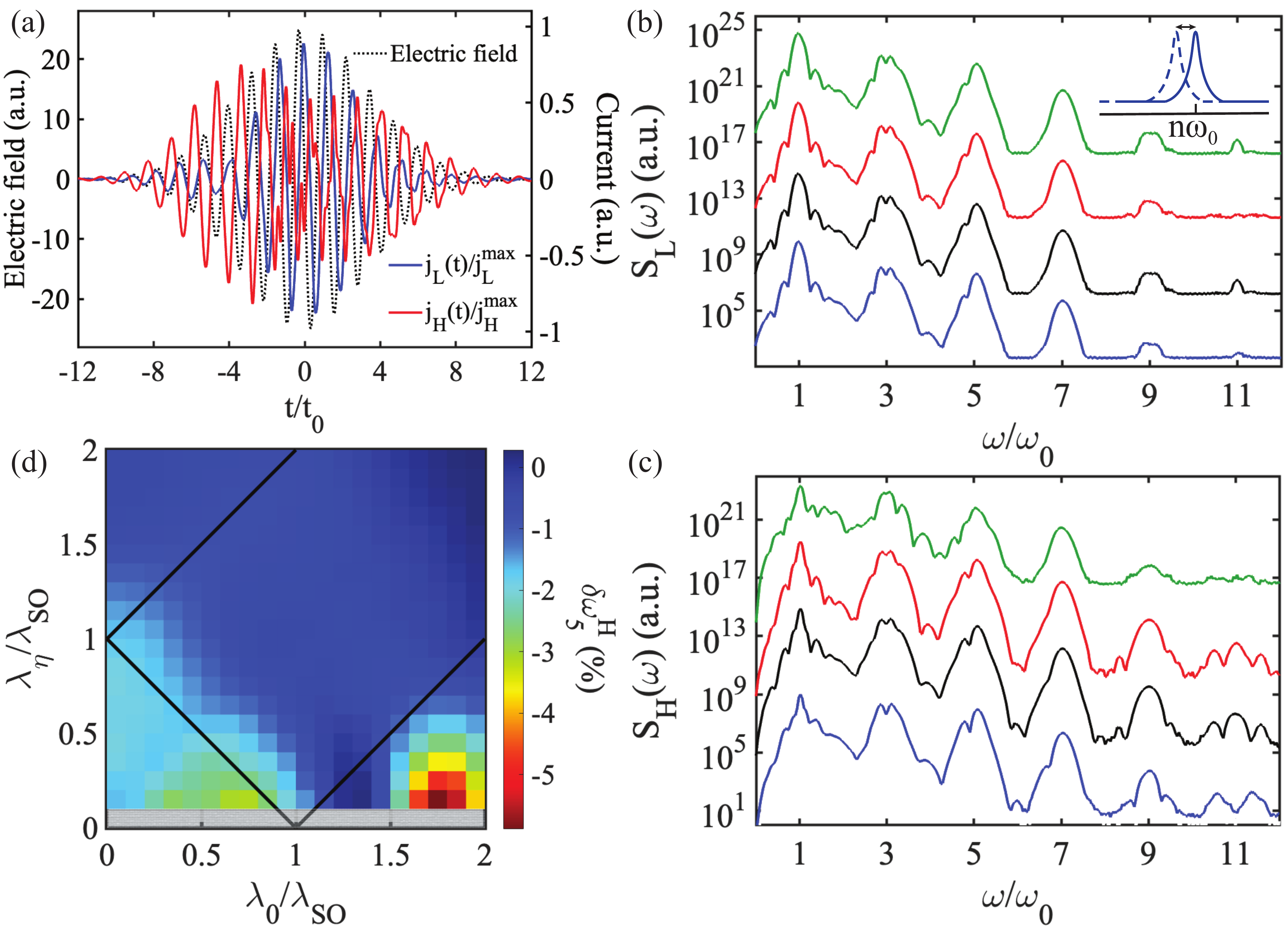}
\caption{{\bf Non-adiabatic quantum dynamics of topological Dirac fermions.} (a) Comparison of $j_L(t)$ and $j_H(t)$ with the driving field highlights the anharmonic  evolution of the currents in the system for the AQHI phase. High harmonic spectrum due to (b) $S_L(\omega)$ and (c) $S_H(\omega)$  for the same electronic phases considered in Fig. \ref{F1}. For visualization purposes we have vertically shifted ($10^5$ a.u.) the curves' baselines.  The inset in (b) is a schematic representation of the relative frequency shift $\delta\omega_n^{L,H} $ of each harmonic with respect to their nominal value $n\omega_0$.  (d) Effects of the topological phase transitions in $\delta\omega_n^{H}$ for the 5th harmonic in the Hall field. Here, $\delta\omega_n^{L,H} = \omega_n^{L,H}/n\omega_1^{L,H} - 1$, where $\omega_n^{L,H} = \int_{(n-0.5)\omega_0}^{(n+0.5)\omega_0} \omega S_{L,H}(\omega) d\omega / \int_{(n-0.5)\omega_0}^{(n+0.5)\omega_0}S_{L,H}(\omega) d\omega$ is the intensity weighed mean emission frequency for the $n$-th harmonic. All parameters are the same as in Fig. \ref{F1}.}
\label{F2}
\end{figure}

We numerically solve the Dirac-Bloch equations and compute the photo-induced currents to unveil the physical mechanisms governing the ultrafast dynamics of topological Dirac-like massive fermions in the monolayer. We focus our discussion to the $(\lambda_0, \lambda_\eta)$ plane, but similar conclusions hold for the entire phase diagram.  Figs. \ref{F1}c,d show the temporal evolution of the longitudinal and Hall currents for selected points in the topological phase diagram. We notice that the current components have very distinct behavior due to the nature of the quantum transitions that drive each of them, as presented in Figs. \ref{F1}e,f.  On the one hand, intraband transitions in the conduction band dominate $j_L(t)$. This leads to a longitudinal current nearly independent of the chosen electronic phase and quasi-symmetric about the time $t=0$, when the incident field reaches peak intensity. In contrast, the Hall current  is  strongly influenced by the choice of $(\lambda_0, \lambda_\eta)$ and its magnitude is enhanced in topological phases with non-zero Chern number ${\cal C}$ since $j_H(t)$ is largely governed by interband transitions.  We  note that the Hall current is excited earlier in the ultrafast process, presents an asymmetric temporal response with respect to the center of the optical pulse, and has a longer duration than its longitudinal counterpart. This is because the weight of contributions from competing transitions varies as the strong field modulates the hot electron population, and we assume that the material is originally in thermal equilibrium with Fermi energy $E_F = 0$. Initially, only interband transitions are possible and lead to the early excitation of Hall currents. As the conduction band gets populated, interband transitions near the band gap become energetically forbidden while intraband transitions are enabled. This results in a small decrease in $j_H(t)$ and excitation of late longitudinal currents.  Finally, as electron-hole pairs relax and recombine, low-energy interband transitions resume, leading to a second peak in the envelope of $j_H(t)$ before the interaction with the optical pulse vanishes. The complex ultrafast electronic response of QSHIs clearly demonstrates the critical role of nonlinear processes in the system. 
\begin{figure}
\includegraphics[width=1\linewidth]{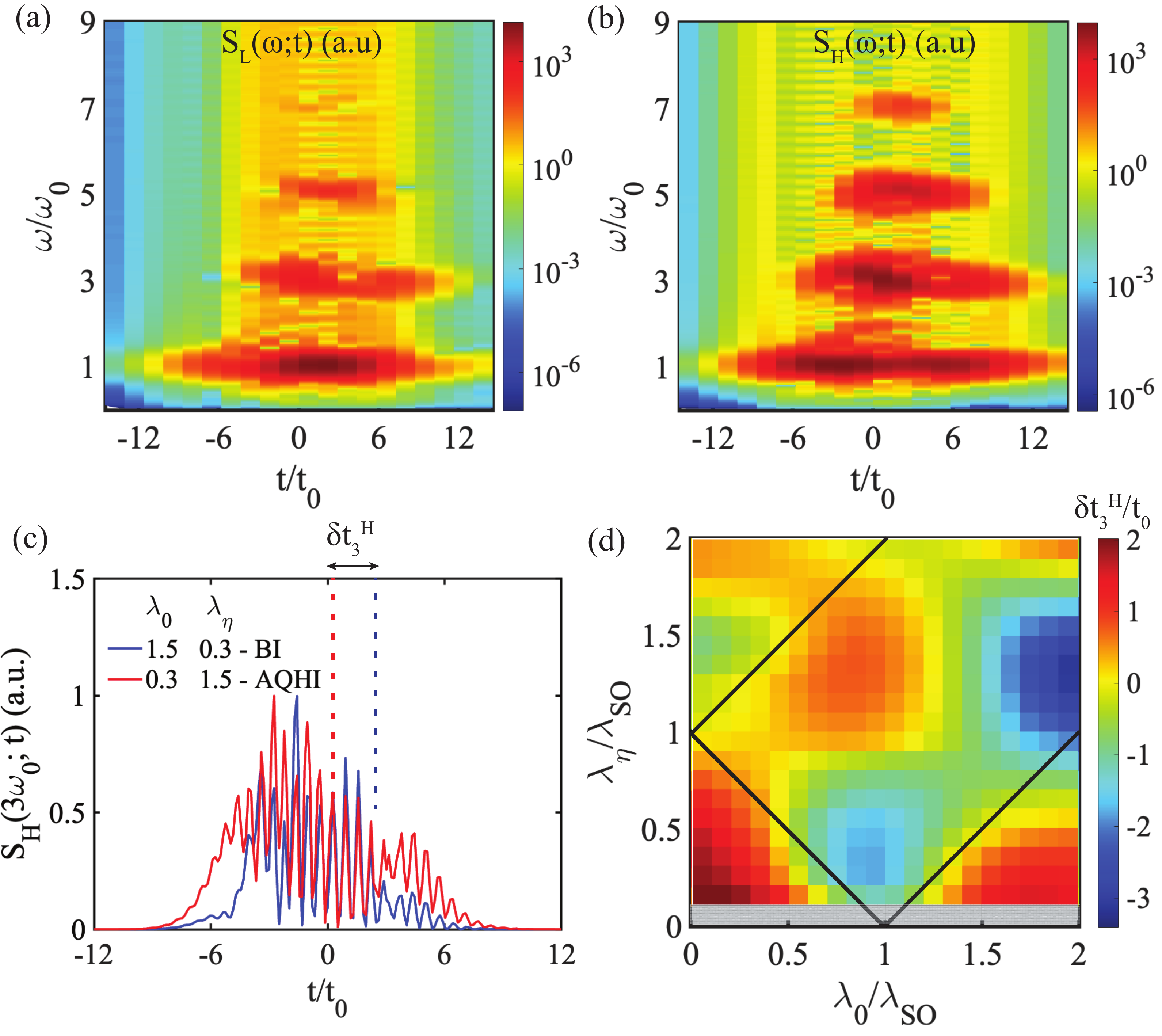}
\caption{{\bf Time-resolved  harmonic emission in quantum spin-Hall insulators.} Short-time Fourier transform $S_L(\omega; t)$  and $S_H(\omega; t)$ of the (a) longitudinal and (b) Hall currents with a Gaussian window (Gabor transform) for the AQHI phase. (c) Dynamics of emission of the 3rd harmonic for two energetically equivalent, but topologically distinct, points in the phase space: $(\lambda_0, \lambda_\eta, \lambda_s) = (0.3, 1.5, 0)$ and $(\lambda_0, \lambda_\eta, \lambda_s) = (1.5, 0.3, 0)$. (d) Average time of emission $\delta t^H_n = \int_{-\infty}^{\infty}  t S_H(\omega_n^{H}; t) dt /\int_{-\infty}^{\infty} S_H(\omega_n^{H}; t) dt$ due to the Hall current for the 3rd harmonic  in the $\lambda_s = 0$ plane of the phase diagram. All parameters are the same as in Fig. \ref{F1}.}
\label{F3}
\end{figure}

In Fig. \ref{F2} we investigate the back-action of the hot electron population on the light pulse. Because the monolayer mass gaps are significantly smaller than the energy of the incident photons, the impinging electromagnetic wave near instantaneously photo-generates free carriers that synchronously follow the electric field oscillations during the leading edge of the pulse (Fig. \ref{F2}a).  We notice that the longitudinal (Hall) current is in-(out-of-) phase with the incident field. The excitation of electron hole-pairs occurs in a time scale faster than the duration of the optical pulse, which results in a switch from a semiconducting to a metal-like material response with a time-dependent plasma frequency  as the interaction evolves. Consequently, the trailing edge of the light pulse envelope probes a transient and rapidly changing electronic population. This causes non-adiabatic quantum evolution of the fermionic currents, which develop a temporal lag with respect to the driving field at later times in the interaction, as seen in Fig. \ref{F2}a. Figures \ref{F2}b,c show that the anharmonic response of the topological Dirac-like fermions results in the generation of scattered fields with higher order odd harmonics of the fundamental frequency. We can clearly distinguish up to the 9th harmonic in the emission spectra $S_{L,H}(\omega) = |\omega j_{L,H}(\omega)|^2$, where $j_{L,H}(\omega)$ are the Fourier transforms of $ j_{L,H}(t)$. The Hall spectra present some signatures of the band structure's intrinsic topology, but the longitudinal ones are quasi-invariant with respect to the electronic phase of the system. The intensities of the harmonics for the phases with ${\cal C} = 0$ are similar, while for the topological phases with ${\cal C}\neq 0$ they are significantly larger, which is consistent with our findings in Fig.~\ref{F2}c,d. Interestingly, we notice that QSHI monolayers emit high harmonics of the incident field at frequencies which are shifted with respect to their nominal value (inset in Fig. \ref{F2}b), similar to observations in graphene \cite{Baudisch2018}. In Fig. \ref{F2}d we show that the underlying topological phase of two-dimensional QHSIs affects the relative frequency shift of $S_H(\omega)$, being enhanced (suppressed) in phases with ${\cal C}=0$ (${\cal C}\neq 0$). 
\begin{figure}
\includegraphics[width=1\linewidth]{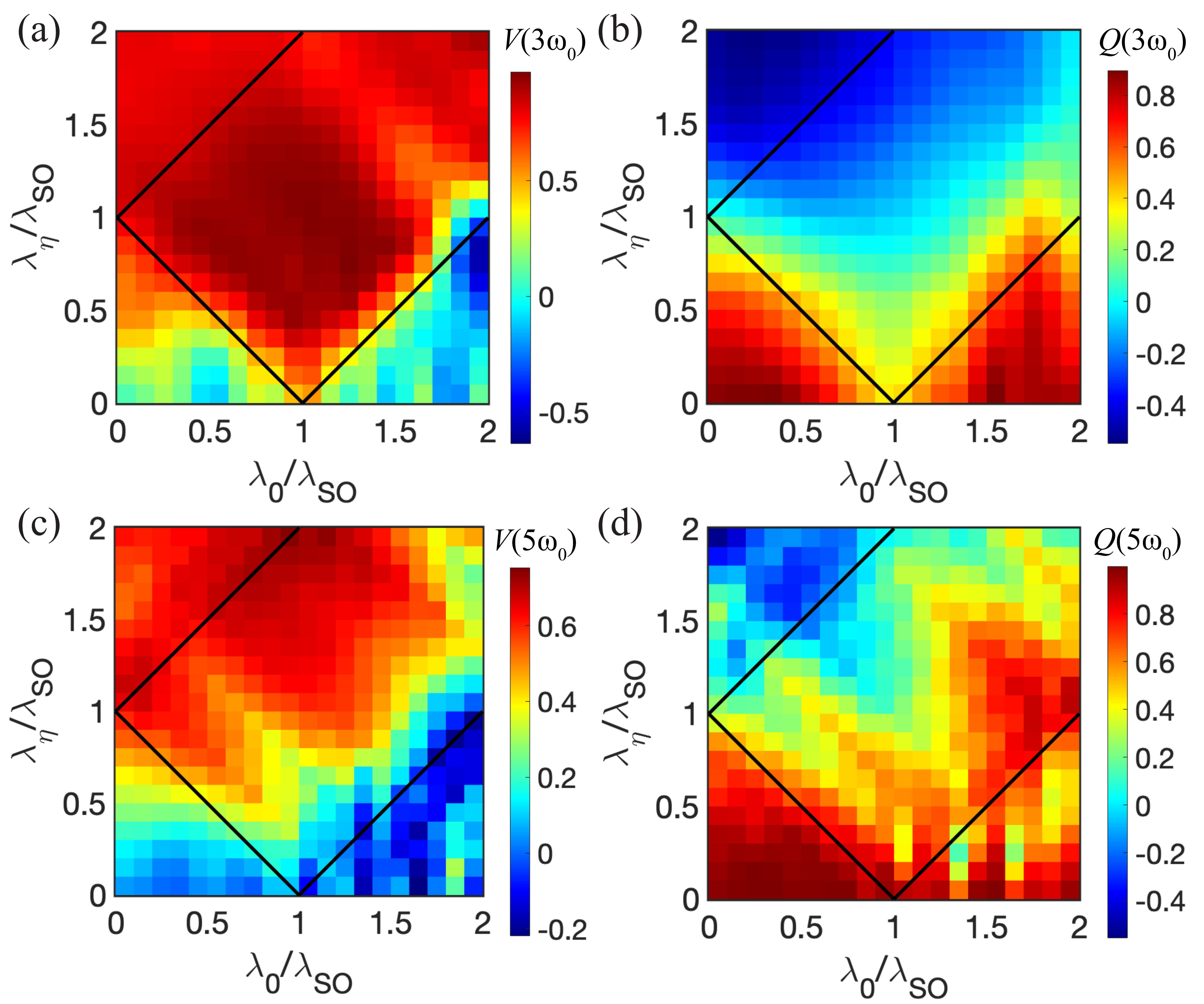}
\caption{{\bf Topology fingerprints in the polarization state.} Phase diagram of the Stokes parameters (a), (c) $V$ for circular and (b), (d) $Q$ for linear polarization for the 3rd (top) and 5th (bottom) harmonics. All parameters are the same as in Fig. \ref{F1}.}
\label{F4}
\end{figure}

In Figs. \ref{F3}a,b we show the time-resolved emission spectrum  of harmonics for a representative  phase with a non-zero Chern number.  The plots reveal that  higher harmonics are more likely to be excited in $S_H(\omega)$  than in $S_L(\omega)$. They show that while the fundamental harmonic is continuously excited,  higher harmonics are generated at later times during the interaction. Our results also demonstrate that the initial time of excitation of a harmonic and its duration scale inversely with the frequency of emission. Thus, HHG is generally confined to time intervals when the incident field approaches its maximum intensity (i.e., $t \simeq 0$), and only lower order harmonics appear near the leading and trailing edges of the optical pulse.  Fig. \ref{F3}c reveals that the dynamics of individual spectral components is affected by the underlying electronic phase of the monolayer. Indeed, Fig. \ref{F3}d shows the mean time of emission $\delta t_3^H$ of the 3rd harmonic of the Hall current, unveiling a serendipitous dependence on the external parameters $\lambda_0$ and $\lambda_\eta$. It evidences that the 3rd harmonic is emitted  with a slight delay with respect to the incident pulse peak power, with larger (smaller) temporal lags occurring for phases with $\mathcal{C} = 0$ ($\mathcal{C} \neq 0$).  Similar results hold for other harmonics (not shown). This topology-dependent time delay in harmonic emission stems from the cross-coupling between the band structure's Berry curvature and the time varying incident field, which offsets  from the Dirac points $K$  and $K'$ the optimum momentum for valence-to-conduction band charge carrier injection, similar to the case of Chern insulators \cite{HHG5}. These findings  indicate the potential of time-resolved harmonic spectroscopy for identifying topological invariants in solid state systems.

In Fig. \ref{F4} we investigate the polarization state of the emitted harmonics, which strongly depends on the intrinsic topology of the energy band structure via the coupling constants $\lambda_{0}$, $\lambda_{\eta}$, and $\lambda_{s}$.  For concreteness we consider the normalized Stokes parameters $V = -2\textrm{Im}[j_{L}(\omega) j_{H}^*(\omega)]/I$ and $Q = (|j_{L}(\omega)|^2 - |j_{H}(\omega)|^2)/I$, where  $I = |j_{L}(\omega)|^2 + |j_{H}(\omega)|^2$ is the frequency-resolved intensity of the field.  Note that $V$ is also referred in the literature as helicity and has been previously employed to distinguish between topologically trivial and nontrivial phases in the Haldane model via circular polarization harmonic emission \cite{HHG5, HHG6, Chacon2020},  while $Q$ represents the asymmetry between harmonics generated with linear polarization parallel to either the longitudinal or the Hall currents. Figs. \ref{F4}a,c  show that the Stokes parameter $V$ varies significantly depending on the electronic phase and clearly distinguishes  phases with zero Chern and non-zero Chern numbers.  The changes in $V$ highlight that in topologically trivial (non-trivial) phases the state of the generated harmonics is primarily dominated by right (left) circular polarization. Note, however, that $V$ fails to differentiate between two non-trivial topological phases. This can be resolved by noticing that the Stokes parameter $Q$ (Figs. \ref{F4}b,d) not only separates  topologically trivial and non-trivial electronic states, but it also enables one to distinguish between phases with non-zero Chern numbers (e.g., AQHI with ${\cal C}=-2$ and PS-QHI with ${\cal C}=-1$).  This follows from the increase of the nonlinear Hall current with the Chern number, hence enhancing the emission of high harmonics polarized orthogonally to the incident light and  providing a mechanism to investigate the topology of the monolayer beyond the linear response regime. 


In summary, we have developed a comprehensive theoretical and numerical framework for investigating topological phase transitions in monolayer topological quantum spin-Hall materials via ultrafast nonlinear photonic processes. As a prototype example, we considered systems described via a generalized Kane-Mele Hamiltonian that accounts for spin-orbit coupling and includes a diversity of knobs that can be controlled externally to drive the system across a multitude of topological phase transitions.  We unveiled the full dynamics of the materials when interacting with strong ultrashort light pulses and showed that various physical quantities can be used to identify and characterize the materials' electronic state.  Recent progress in the synthesis of various topological semiconductors, e.g.,  graphene family monolayers and Jacutingaite materials, together with advances in  nonlinear characterization photonic techniques implies that  our results can be accessed experimentally with current technologies. For example, measurement of Stokes parameters can be realized by employing commercially available polarizers. The frequency shift and time delay are  $\sim 10-50$ THz and $\sim 1-10$ fs for the 3rd harmonic when we consider an incident laser pulse with intensity $3000$ GW/cm$^2$, $\omega_0/2\pi = 360$ THz, $\tau = 30$ fs and a monolayer with $\lambda_{SO} = 0.3$ meV, all within existing measurement capabilities.  We envision that our findings will impact the research at the intersection of  ultrafast optics, topological materials, spintronics, and valleytronics.  

\begin{acknowledgments} 
R.K.M and W.K.K.  acknowledge the Laboratory Directed Research and Development program of Los Alamos National Laboratory under Project No. 20190574ECR. R.K.M. also thanks the Center for Nonlinear Studies at LANL for financial support under Project No. 20190495CR. R.K.M. was partially supported by the U.S. Department of Energy, Office of Science, Basic Energy Sciences, Materials Sciences and Engineering Division, Condensed Matter Theory Program. DK, AC and DEK are grateful for the support from the National Research Foundation of Korea (NRF) Grants (Grant No. 2022M3H4A1A04074153 and No. 2020R1A2C2103181) funded by the Ministry of Science, ICT, and the Competency Development Program (Grant No. P0008763) through the Korea Institute for Advancement of Technology (KIAT) funded by the Korea Government (MOTIE).
\end{acknowledgments}

\end{document}


\title{Supplementary Material }

\author{Rajesh K. Malla}
\affiliation{Theoretical Division, MS B262, Los Alamos National Laboratory, Los Alamos, New Mexico 87545, USA}
\affiliation{Center for Nonlinear Studies, MS B258, Los Alamos National Laboratory, Los Alamos, New Mexico 87545, USA}
\author{Dasol Kim}
\affiliation{Department of Physics and Center for Attosecond Science and Technology, POSTECH, 7 Pohang 37673, South Korea}
\affiliation{Max Planck POSTECH/KOREA Research Initiative, Pohang 37673, South Korea}
\author{Dong Eon Kim}
\affiliation{Department of Physics and Center for Attosecond Science and Technology, POSTECH, 7 Pohang 37673, South Korea}
\affiliation{Max Planck POSTECH/KOREA Research Initiative, Pohang 37673, South Korea}
\author{Alexis Chac\'{o}n}
\affiliation{Department of Physics and Center for Attosecond Science and Technology, POSTECH, 7 Pohang 37673, South Korea}
\affiliation{Max Planck POSTECH/KOREA Research Initiative, Pohang 37673, South Korea}
\author{Wilton J. M. Kort-Kamp}
\email{kortkamp@lanl.gov}
\affiliation{Theoretical Division, MS B262, Los Alamos National Laboratory, Los Alamos, New Mexico 87545, USA}

\maketitle

\section{Derivation of ultrafast current\label{sec:1}}
The wavefunction of a massive Dirac fermion  satisfies the Schr{\" o}dinger equation 
\begin{equation}
i\hbar\frac{\partial}{\partial t} |\psi_{{\bf k}}^{\eta,s}(t)\rangle=H_{s}^{\eta}({\bf k},t)|\psi_{{\bf k}}^{\eta,s}(t)\rangle,
    \label{dirac}
\end{equation}
where ${\textbf{p}}=  \hbar {\bf k} = \hbar(k_x,k_y)$ is the particle momentum, $v_F$ is the Fermi velocity, $\tau_{i}$ are the sub-lattice pseudo-spin Pauli matrices, and $\eta, s = \pm 1$ are valley and spin indexes. Here, the time-dependent Hamiltonian $H_{s}^{\eta}({\bm k},t)$ is obtained by shifting the particle momentum $\hbar{\bf k}\rightarrow \boldsymbol{\Pi}_{\textbf{k}}(t)   =\hbar{\bf k}- e{\bm A}(t)$ in the static Hamiltonian ${H}_s^{\eta}({\bf k})= \hbar v_F(\eta k_x \tau_x +k_y \tau_y) + \Delta_s^{\eta} \tau_z$ as defined in the main text. ${\bm A}(t)$ is the vector potential of the normally incident electric field.

The general form of the wave function of a fermion in a Dirac cone $(\eta, s)$ can be expressed as 
\begin{equation}
|\psi_{{\bf k}}^{\eta,s}(t)\rangle=\sum\limits_{\lambda=\pm 1}C_{\lambda {\bf k}}^{\eta,s}(t) |\psi_{\lambda{\bf k}}^{\eta,s}(t)\rangle
    \label{lin-comb},
\end{equation}
where $\lambda=+1 (-1)$ corresponds to the conduction (valence) band, and the time-dependent wave functions $|\psi_{\lambda{\bf k}}^{\eta,s}(t)\rangle$ can be expressed as
\begin{equation}
|\psi_{\lambda {\bf k}}^{\eta,s}(t)\rangle=|u_{\lambda {\bf k}}^{\eta,s}(t)\rangle \times \exp{\left\{ \dfrac{-i }{\hbar}\displaystyle{\int_{-\infty}^{t}} dt' \left[{E_{\lambda\bf k}^{\eta,s}(t')}-\lambda \hbar \dot{\gamma}_{{\bf k}}^{\eta,s}(t')\right]\right\}},
    \label{psi-1}
\end{equation}
where $|u_{\lambda {\bf k}}^{\eta,s}(t)\rangle$ are the instantenous eigenvectors of the Hamiltonian $H_{s}^{\eta}({\bf k},t)$, which satisfy 
\begin{equation}
H_{s}^{\eta}({\bf k},t)|u_{\lambda {\bf k}}^{\eta,s}(t)\rangle=E_{\lambda{\bf k}}^{\eta,s}(t)|u_{\lambda {\bf k}}^{\eta,s}(t)\rangle,
\end{equation}
and the the orthogonality relation 
$\langle u_{\lambda {\bf k}}^{\eta,s}(t)|u_{\lambda' {\bf k}}^{\eta',s'}(t)\rangle=\delta_{\lambda\lambda'}\delta_{\eta\eta'}\delta_{ss'}.  
$
The eigenenergies and the eigenvectors are
\begin{equation}
    E_{\lambda{\bf k}}^{\eta,s}(t)=\lambda \epsilon_{{\bf k}}^{\eta,s}(t)=\lambda \left( v_F^2  {\Pi}_{\textbf{k}}(t) ^2+{\Delta_{s}^{\eta}}^2\right)^{1/2},
    \label{eig-energy}
\end{equation}
and 
\begin{equation}
    |u_{\lambda {\bf k}}^{\eta,s}(t)\rangle=\frac{v_F{\Pi}_{\textbf{k}}(t) }{\Big[2\epsilon_{{\bf k}}^{\eta,s}(t)(\epsilon_{{\bf k}}^{\eta,s}(t)+\lambda \Delta_{s}^{\eta})\Big]^{1/2}}\times\begin{pmatrix}
    \frac{\epsilon_{{\bf k}}^{\eta,s}(t)+\lambda \Delta_{s}^{\eta}}{\eta v_F{\Pi}_{\textbf{k}}(t) } e^{-i \eta \theta_{\textbf{k}}(t)} \\ \\
    \lambda
    \end{pmatrix},
    \label{eig-vector}
\end{equation}
where ${\Pi}_{\textbf{k}}(t)  =| \boldsymbol{\Pi}_{\textbf{k}}(t)| = \sqrt{[\hbar k_x + e A_x(t)]^2 + \hbar k_y^2}$ for a field linearly polarized along the $x-$direction. Also,  $\theta_{{\bf k}}(t)=\tan^{-1} (\Pi_{\textbf{k}_y}(t)/\Pi_{\textbf{k}_x}(t))$, and  $\lambda\dot{\gamma}_{{\bf k}}^{\eta,s}(t)$ $=$ $i\langle u_{\lambda {\bf k}}^{\eta,s}(t)|\frac{\partial}{\partial_{t}}|u_{\lambda {\bf k}}^{\eta,s}(t)\rangle$ $=$ $\eta \lambda\Delta_{s}^{\eta}\dot{\theta}_{{\bf k}}(t)/2\epsilon_{{\bf k}}^{\eta,s}(t)$ is time the derivative of the Berry phase of the corresponding eigenstate and remains nonzero for a Dirac cone with a finite energy band gap. This phase plays a crucial role in introducing topological signatures into the nonlinear dynamics, and for the incident field in consideration, one can show that $\gamma_{{\bf k}}^{\eta,s}(t) = [\Lambda_{{\bf k}}^{\eta,s}(t) - \Lambda_{{\bf k}}^{\eta,s}(-\infty)]/4$, where 
\begin{equation}
 \Lambda_{{\bf k}}^{\eta,s}(t) = \textrm{tan}^{-1} \left[\dfrac{2\Delta_{s}^{\eta}\epsilon_{{\bf k}}^{\eta,s}(t) \tan \theta_{\textbf{k}}(t)}
 {{\Delta_{s}^{\eta}}^2-\epsilon_{{\bf k}}^{\eta,s}(t)^2 \tan^2 \theta_{\textbf{k}}(t)}\right].
\end{equation}

Substituting (\ref{psi-1}) into (\ref{lin-comb}), and then (\ref{lin-comb}) into (\ref{dirac}) and using the orthonormality relation,  one can derive coupled differential equations for the complex amplitudes $C_{\lambda {\bf k}}^{\eta,s}(t)$. It is convenient to introduce the population inversion $w_{{\bf k}}^{\eta,s}(t)\equiv|C_{+1{\bf k}}^{\eta,s}(t)|^2-|C_{-1{\bf k}}^{\eta,s}(t)|^2$ and the interband coherence $\rho_{{\bf k}}^{\eta,s}(t)\equiv C_{+1 {\bf k}}^{\eta,s}(t)C_{-1 {\bf k}}^{\eta,s^*}(t)\exp\left[-\frac{2i}{\hbar}\int\limits_{-\infty}^{t} dt' \left(\epsilon_{\bf k}^{\eta,s}(t')-\eta\hbar \dot{\gamma}_{{\bf k}}^{\eta,s}(t')\right)\right],$ which are physically more meaningful. These quantities satisfy the following coupled differential equations
\begin{equation}
    \dot{w}_{{\bf k}}^{\eta,s}(t)=-\rho_{{\bf k}}^{\eta,s}(t)\left(\frac{i\eta v_F  \Pi_{\bf k}(t)\dot{\theta}_{{\bf k}}(t)}{\epsilon_{{\bf k}}^{\eta,s}(t)} +\frac{\Delta_s^{\eta} v_F \dot{ \Pi}_{{\bf k}}(t)}{{\epsilon_{{\bf k}}^{\eta,s}}(t)^2}\right)+c.c.
    \label{d1}
\end{equation}
\begin{equation}
    \dot{\rho}_{{\bf k}}^{\eta,s}(t)=-2i\rho_{{\bf k}}^{\eta,s}(t)\left(\frac{\epsilon_{{\bf k}}^{\eta,s}(t)}{\hbar}-\dot{\gamma}_{{\bf k}}^{\eta,s}(t)\right)\\
-w_{{\bf k}}^{\eta,s}(t)\left(\frac{i\eta v_F  \Pi_{\bf k}(t)\dot{\theta}_{{\bf k}}(t)}{2\epsilon_{{\bf k}}^{\eta,s}(t)} -\frac{\Delta_s^{\eta} v_F \dot{ \Pi}_{{\bf k}}(t)}{2{\epsilon_{{\bf k}}^{\eta,s}}(t)^ 2}\right),
\label{d2}
\end{equation}
 known as Dirac-Bloch equation. Assuming that at equilibrium the monolayer is neutral with Fermi energy equal to zero, the population and coherence should satisfy the following initial conditions in the limit of vanishing temperature $w_{{\bf k}}^{\eta,s}(t \rightarrow -\infty) = -1$ and $\rho_{{\bf k}}^{\eta,s}(t \rightarrow -\infty) = 0$.

The ultrafast nonlinear electronic response of the monolayer can be analyzed viathe time-dependent electric current ${\bm j}(t)= j_L(t)\hat{\textbf{x}}+ j_H(t)\hat{\textbf{y}}$. 
The current density operator operator for longitudinal and Hall currents, respectively, are  ${\hat j}_{L,{\bf k}}=-(e/\hbar)(\partial H_{{\bf k}}^{\eta,s}/\partial k_{x})$ and ${\hat j}_{H,{\bf k}}=-(e/\hbar)(\partial H_{{\bf k}}^{\eta,s}/\partial k_{y})$. The momentum-dependent current is then given by ${\bm j}_{\nu,{\bf k}}^{\eta,s}(t)=\langle \psi_{{\bf k}}^{\eta,s}(t)|{\hat j}_{\nu,{\bf k}}| \psi_{{\bf k}}^{\eta,s}(t) \rangle-\langle \psi_{-1{\bf k}}^{\eta,s}(t)|{\hat j}_{\nu,{\bf k}}| \psi_{-1{\bf k}}^{\eta,s} (t)\rangle$, where $\nu = L, H$ \cite{Baudisch2018}. The second term is considered as an ad hoc subtraction of valence band generated current. This eliminates the unphysical divergences that appears due to the first term which is an artifact of the low energy dispersion. 
This further eliminates the dissipative current in the valence band which is absent in the tight-binding model \cite{sipe1993}.
In terms of the population inversion and interband coherence, the momentum dependent current density can be expressed as
\begin{widetext}
\begin{equation}
{\bm j}_{{\bf k}}^{\eta,s}(t)=-e v_F \begin{pmatrix}
\eta \cos \theta_{{\bf k}}(t) && \eta \sin \theta_{{\bf k}}(t)\\ \\\sin \theta_{{\bf k}}(t) && -\cos \theta_{{\bf k}}(t)
\end{pmatrix} \begin{pmatrix}
\frac{v_F\Pi_{{\bf k}}(t)(w_{{\bf k}}^{\eta,s}(t)+1)}{\epsilon_{{\bf k}}^{\eta,s}(t)}-\frac{2\Delta_s^{\eta}}{\epsilon_{{\bf k}}^{\eta,s}(t)} \text{Re}[\rho_{{\bf k}}^{\eta,s}(t)]\\\\
-2\text{Im}[\rho_{{\bf k}}^{\eta,s}(t)]
\end{pmatrix}.
    \label{current}
\end{equation}
\end{widetext}
The terms proportional to the population ($w_{{\bm k}}^{\eta,s}$) represent the contributions from intraband transitions  to the current, while the remaining terms (proportional to polarization $\rho_{{\bm k}}^{\eta,s}$) represent the contributions arising from interband to the current. Finally, the total time current density is obtained by integrating the above expression over the entire momentum space and adding the contributions from all Diract cones,
\begin{equation}
\textbf{j}(t) = \dfrac{1}{4\pi^2} \sum_{\eta, s}\int d\textbf{k}\ {\bm j}_{{\bf k}}^{\eta,s}(t),
\label{int-current}
\end{equation} 
while the spectrum of harmonic emission in the systems follows by Fourier transforming  the above expression to the frequency domain.

\section{Universal Dirac-Bloch Equations in Two-dimensional Kane-Mele QSHI materials}
To investigate the dynamics of various two-dimensional QSHI materials, one requires solving  (\ref{d1}) and (\ref{d2}) many times with different material properties, {\it i.e.,} intrinsic spin-orbit coupling strength $\lambda_{SO}$. These computations can be computationally demanding and time-consuming. Fortunately, universal equations modeling the nonlinear ultrafast dynamics of \textit{all} 2D materials described by the Kane-Mele Hamiltonian with a non-zero spin-orbit coupling can be derived by expressing all physical quantities units of the fundamental energy scale of the system, namely,  $\lambda_{SO}$. One can express the material properties such as energy, momentum, and Dirac mass gap in terms of dimensionless variables (denoted with a tilde) as
\begin{equation}
\epsilon_{\textbf{k}}^ {\eta,s} (t)={\tilde \epsilon}_{\textbf{k}}^ {\eta,s} (t) \lambda_{SO}, 
\boldsymbol{\Pi}_{\textbf{k}}(t)=\tilde{\boldsymbol{\Pi}}_{\textbf{k}}(t) \frac{\lambda_{SO}}{v_{F}}, 
\Delta^ {\eta}_s={\tilde \Delta}^ {\eta}_s\lambda_{SO}.
\label{var-change1}
\end{equation}
Similarly, the parameters of the incident optical field can be put as
\begin{equation}
A_0= {\tilde A}_0 \frac{ \lambda_{SO}}{e v_F}, 
\omega_0=\tilde{\omega}_0 \frac{ \lambda_{SO}}{\hbar},
\tau = \tilde{\tau} \frac{\hbar}{\lambda_{SO}}.
\label{var-change2}
\end{equation}
By defining $t_0 = \hbar/\lambda_{SO}$ as the fundamental temporal scale, expressing the time variable in dimensionless units as $t = \tilde{t} t_0$, and noticing that any function of time $f(t)$ can be written as $f(t) = f(\tilde{t}t_0) = : \tilde{f}(\tilde{t})$, one can show that $\dot{f}(t) \rightarrow t_0^ {-1} \dot{\tilde{f}}(\tilde{t})$. As a result of these changes of variables  Eqs. \eqref{d1} and \eqref{d2} can be cast in a universal form applicable to any two-dimensional Kane-Mele QSHI material, regardless of specific values of the spin-orbit coupling: 
\begin{equation}
    \dot{{\tilde w}}_{{\bf {\tilde k}}}^{\eta,s}({\tilde t})=-\tilde{\rho}_{{\bf \tilde{k}}}^{\eta,s}(\tilde{t})\left(\frac{i\eta  \tilde{\Pi}_{\bf \tilde{k}}(\tilde{t})\dot{\tilde{\theta}}_{{\bf \tilde{k}}}(\tilde{t})}{\tilde{\epsilon}_{{\bf \tilde{k}}}^{\eta,s}(\tilde{t})} +\frac{\tilde{\Delta}_s^{\eta} \dot{ \tilde{\Pi}}_{{\bf \tilde{k}}}(\tilde{t})}{{{{\tilde \epsilon}_{{\bf \tilde{k}}}^{\eta,s}}}(\tilde{t})^ 2}\right)+c.c.
    \label{d1tilde}
\end{equation}
\begin{equation}
    \dot{\tilde{\rho}}_{{\bf \tilde{k}}}^{\eta,s}(\tilde{t})=-2i\tilde{\rho}_{{\bf \tilde{k}}}^{\eta,s}(t)\left(\tilde{\epsilon}_{{\bf \tilde{k}}}^{\eta,s}(\tilde{t})-\dot{\tilde{\gamma}}_{{\bf \tilde{k}}}^{\eta,s}(\tilde{t})\right)\\
-\tilde{w}_{{\bf \tilde{k}}}^{\eta,s}(\tilde{t})\left(\frac{i\eta  \tilde{\Pi}_{\bf \tilde{k}}(\tilde{t})\dot{\tilde{\theta}}_{{\bf \tilde{k}}}(\tilde{t})}{2\tilde{\epsilon}_{{\bf \tilde{k}}}^{\eta,s}(\tilde{t})} -\frac{\tilde{\Delta}_s^{\eta}  \dot{ \tilde{\Pi}}_{{\bf \tilde{k}}}(\tilde{t})}{{2\tilde{\epsilon}_{{\bf \tilde{k}}}^{\eta,s}}(\tilde{t})^2}\right).
\label{d2tilde}
\end{equation}
The solution of Eqs. (\ref{d1tilde}) and (\ref{d2tilde}) represents dimensionless population and interband-coherence and can be used to compute universal  dimensionless microscopic currents ${\bm \tilde{j}}_{{\bf \tilde{k}}}^{\eta,s}(\tilde{t})$ from Eq. (\ref{current}) and ${\bf {\tilde j}}(\tilde{t})$ from Eq.  (\ref{int-current}) in a similar fashion to the above derivation. Note that \textit{all} the dimensionless quantities are expressed in the units of $\lambda_{SO}$. Therefore, the solutions of Eqs. (\ref{d1tilde}) and (\ref{d2tilde}) can be extended to all QSHI materials for fixed parameters of the incident field,  where one can obtain all the  quantities in physical units rescaled by a a single dimensionfull quantity $\lambda_{SO}$.
\section{Numerical evaluation of ultrafast current\label{sec:2}}
The microscopic current is computed using equation (\ref{current}), where the population inversion and interband coherence satisfy the coupled differential equations (\ref{d1}) and (\ref{d2}). We use an ordinary differential equation (ODE) solver (ODE45 in MATLAB) using the Runge-Kutta(4,5) method to solve the coupled differential equations. The dimensionless time interval is $0.01$ and the absolute tolerance for the ODE45 is $10^{-8}$.   The time-dependent total current ${\bm j}(t)$ is obtained by numerically integrating the microscopic currents as shown in equation (\ref{int-current}). The numerical integration is performed over a  polar grid with $10^6$ points of evaluation in momentum space. A radial cutoff is imposed at in the radial direction  in order to include all the relevant microscopic contributions.

